\documentclass[12pt,preprint]{aastex}
\usepackage{epsfig}

\begin{document}
\title{~~\\ ~~\\ A shrinking Compact Symmetric Object: J11584+2450?}
\shorttitle{CSO J11584+2450}
\author{ S. E. Tremblay\altaffilmark{1},  G. B. Taylor\altaffilmark{1,2},  J. F. Helmboldt\altaffilmark{3}, C. D. Fassnacht\altaffilmark{4}, and T. J. Pearson\altaffilmark{5} }
%\email{gtaylor@nrao.edu}

\altaffiltext{1}{Department of Physics and Astronomy, University of New Mexico, Albuquerque, NM 87131; tremblay@unm.edu, gbtaylor@unm.edu}
\altaffiltext{2}{National Radio Astronomy Observatory, Socorro NM 87801}
\altaffiltext{3}{Naval Research Laboratory, Code 7213, Washington, DC 20375; joe.helmboldt@nrl.navy.mil}
\altaffiltext{4}{Department of Physics, University of California, Davis, CA 95616; fassnacht@solid.physics.ucdavis.edu}
\altaffiltext{5}{Astronomy Department, California Institute of Technology, Pasadena, CA 91125; tjp@astro.caltech.edu}
%%%%%%%%%%%%%%%%%%%%%%%%%%%%%%%%%%%%%%%%%%%%%%%%%%%%%%%%%%%%%%%%%%%%%%%%%%%%%%%%%%%%%%%%%%%
%                               ABSTRACT                                                  %
%%%%%%%%%%%%%%%%%%%%%%%%%%%%%%%%%%%%%%%%%%%%%%%%%%%%%%%%%%%%%%%%%%%%%%%%%%%%%%%%%%%%%%%%%%%

\begin{abstract}

We present multi-frequency multi-epoch Very Long Baseline Array (VLBA) observations of J11584+2450.   These observations clearly show this source, previously classified as a core-jet, to be a compact symmetric object (CSO). Comparisons between these new data and data taken over the last 9 years shows the edge brightened hot spots retreating towards the core (and slightly to the west) at approximately 0.3c. Whether this motion is strictly apparent or actually physical in nature is discussed, as well as possible explanations, and what implications a physical contraction of J11584+2450 would have for current CSO models.

\end{abstract}


%%%%%%%%%%%%%%%%%%%%%%%%%%%%%%%%%%%%%%%%%%%%%%%%%%%%%%%%%%%%%%%%%%%%%%%%%%%%%%%%%%%%%%%%%%%
%                               INTRODUCTION                                              %
%%%%%%%%%%%%%%%%%%%%%%%%%%%%%%%%%%%%%%%%%%%%%%%%%%%%%%%%%%%%%%%%%%%%%%%%%%%%%%%%%%%%%%%%%%%

\section{Introduction}

Compact symmetric objects (CSOs) are now a well established class of radio sources loosely defined as   sources with emission on both sides of the core (which itself is not always detected) on a size scale of 1 kpc or less \citep{1994ApJ...432L..87W}. The generally accepted explanation for the small size of these objects is that they are young radio sources which could grow into larger FR II objects \citep{1996ApJ...460..612R, 1998PASP..110..493O}. Alternately, it has been proposed that the small size of these structures is due to their growth being frustrated by a dense environment \citep{1984AJ.....89....5V, 1998PASP..110..493O}.

Due to their rapid growth, age estimates for the emission from these objects can be obtained kinematically \citep{1998A&A...337...69O, 2000ApJ...541..112T} yielding ages ranging from tens to thousands of years. Less accurate spectroscopic models \citep{1996ApJ...460..612R, 1999A&A...345..769M, 2005ApJ...622..136G} place CSOs at a few thousands of years old. This generally supports the theory that these are young AGN in the early stages of evolution. However, since the age distribution of currently known CSOs is heavily weighted on the younger side \citep{2005ApJ...622..136G}, this indicates the evolution might not be straightforward. This distribution should not be taken as definitive though, since Gugliucci et al. concede it may be influenced by selection effects. One theory suggests CSOs could be generally short-lived objects with only a small fraction of them surviving to become larger-scale objects while the remaining galaxies become permanently radio quiet \citep{1998A&A...337...69O}.  In a competing theory there exists a cyclic process where unsuccessful CSOs have multiple opportunities to grow into larger objects \citep{1997AJ....113..148O}. Alternatively, the current distribution could be an artifact of the small statistical sample from which it is derived combined with selection effects. 

The picture presented in the above models might be overly simplistic. For example, all of these models predict continuous radial expansion of the lobes, but CSOs such as 1031+567 \citep{2000ApJ...541..112T} have been observed with non-radial motion. Here we present observations of J11584+2450 (PKS 1155+251, SDSS J115825.79+245018.0), a galaxy with a redshift of $0.20160 \pm 0.00040$ \citep{2002AJ....124..662Z} that appears to be contracting towards its core on both sides.

Throughout this discussion, we assume H$_{0}$=73 km s$^{-1}$
Mpc$^{-1}$, $\Omega_m$ = 0.27, $\Omega_\Lambda$ = 0.73, so 1 mas = 3.213 pc.

%%%%%%%%%%%%%%%%%%%%%%%%%%%%%%%%%%%%%%%%%%%%%%%%%%%%%%%%%%%%%%%%%%%%%%%%%%%%%%%%%%%%%%%%%%%
%                               OBSERVATIONS                                              %
%%%%%%%%%%%%%%%%%%%%%%%%%%%%%%%%%%%%%%%%%%%%%%%%%%%%%%%%%%%%%%%%%%%%%%%%%%%%%%%%%%%%%%%%%%%

\section{Observations and Data Reduction}
\label{observations}

Multi-frequency observations of J11584+2450 were performed on September 19, 2006 with the VLBA.  A summary of these and other observations referred to in this paper is presented in Table \ref{Observations}. These observations consisted of four 8 MHz wide IFs in the C, X and U bands with full polarization centered at: 4605.5 , 4675.5, 4990.5, 5091.5, 8106.0, 8176.0, 8491.0, 8590.0, 14902.5, 14910.5, 15356.5 and 15364.5 MHz at an aggregate bit rate of 256 Mbps to maximize (u,v) coverage and sensitivity. When the data in each band were combined, the three central frequencies were: 4844.7, 8344.7, and 15137.5 MHz. The integrations were performed in blocks ($\sim$2 minutes for 5 and 8 GHz, $\sim$7.5 minutes for 15 GHz) and these blocks were spread out over a 9.5 hour period to maximize (u,v) coverage of the source.

Most of the calibration and initial imaging of the new data were carried out by automated AIPS \citep{2003ASSL..285..109G} and DIFMAP \citep{1997ASPC..125...77S} scripts similar to those used in reducing the VIPS 5 GHz survey data \citep{2007ApJ...658..203H, 2005ApJS..159...27T}. To summarize, flagging of bad data and calibration were performed using the VLBA data calibration pipeline \citep{2005ASPC..340..613S}, while imaging was performed using DIFMAP scripts described in \citet{2005ApJS..159...27T}. Final imaging was performed manually using the DIFMAP program, with beam sizes of $1.906\times3.16$ in position angle -6.89$^\circ$, $1.195\times1.788$ in position angle -3.58$^\circ$ and $0.6876\times0.9794$ in position angle -2.717$^\circ$ for 5, 8 and 15 GHz respectively.

%%%%%%%%%%%%%%%%%%%%%%%%%%%%%%%%%%%%%%%%%%%%%%%%%%%%%%%%%%%%%%%%%%%%%%%%%%%%%%%%%%%%%%%%%%%
%                               RESULTS                                                   %
%%%%%%%%%%%%%%%%%%%%%%%%%%%%%%%%%%%%%%%%%%%%%%%%%%%%%%%%%%%%%%%%%%%%%%%%%%%%%%%%%%%%%%%%%%%

\section{Results}
\label{results}

\subsection{Images}
\citet{2004ApJ...609..539K} observed J11584+2450 (B1155+251) as part of the VLBA 2cm Survey. Since these observations were only at 15 GHz, they typically identified the brightest component in an image to be the core, and consequently classified this source as a core-jet with the core being the southern, bright component.

\begin{figure}
\epsscale{0.5}
\plotone{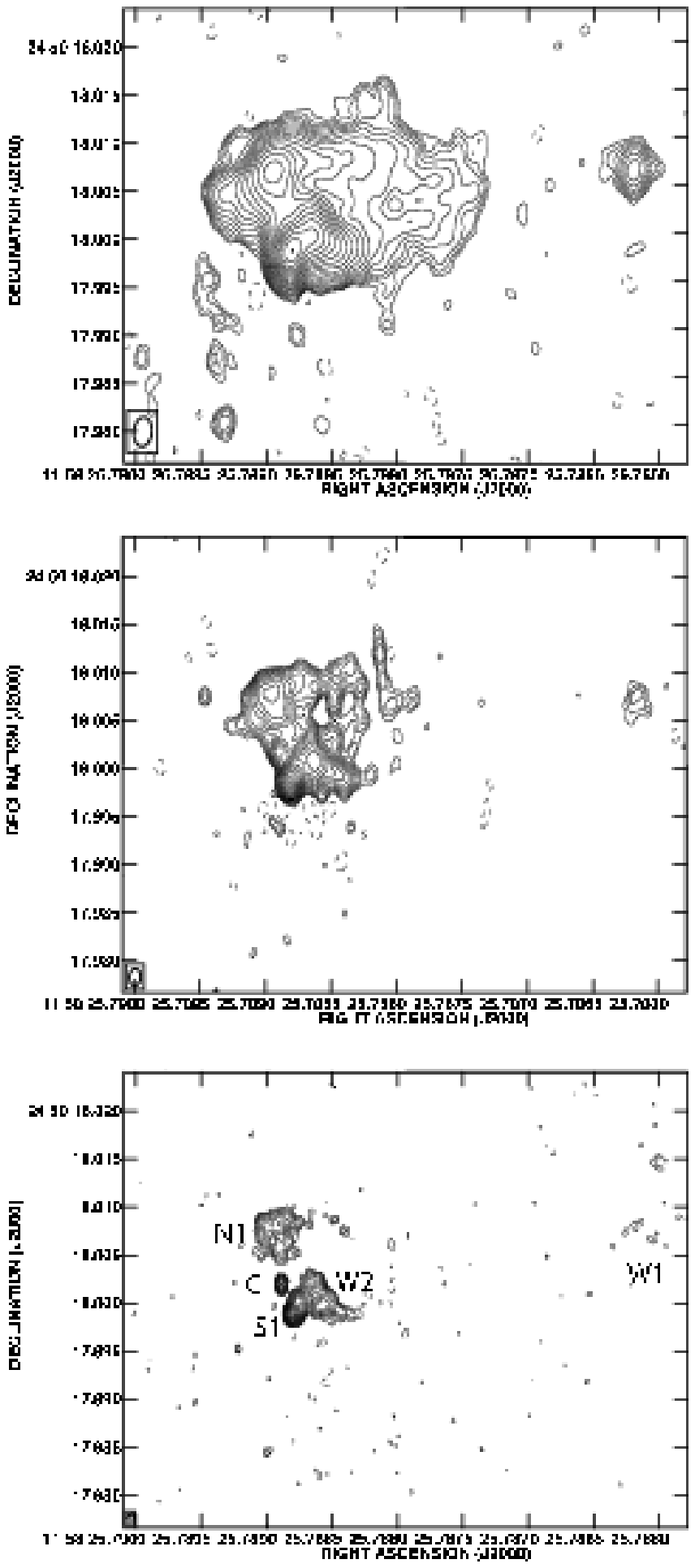}
\caption{VLBA observations from September 2006 of J11584+2450 at frequencies of (moving top to bottom) 4.84, 8.34, and 15.13 GHz. Contour levels begin at 0.375 mJy/beam and increase by factors of $2^{1/2}$.  }
\label{VLBA}
\end{figure}

Figure \ref{VLBA} shows the 5, 8, and 15 GHz VLBA images made from the September 2006 observations of J11584+2450. The 15 GHz map shows the clearest structure so is used to label components of the source. The 15 GHz image shows a compact unresolved component (C) with resolved emission both to the north (N1) and the south (S1). The southern emission then seems to have another component that expands out towards the west (W2). There also exists some weak emission on the western edge of the image (W1; with a peak flux density of 0.55 mJy/beam). S1 is the brightest component in the image (52.3 mJy/beam peak), and is what was previously identified as the core.  In the 8 GHz image the edges of C, N1, S1, and W2 become indistinguishable, but these components can still be identified by local peaks within the image. Interestingly, an eastern spur develops from the southern edge of N1, extending in the opposite direction from the majority of the diffuse emission. Overall the emission appears more extended, and W1 has a more significant detection. The 5 GHz map further smears the interior components together until only N1 and S1 are clearly visible as local maxima of the map. The spur mentioned above becomes brighter, and the western emission stretches out farther towards a very well detected W1 (3.83 mJy/beam peak flux density) . There is emission from the eastern spur towards the south in this image (hereafter referred to as the 5 GHz southeastern clump), which has an integrated flux density of 3.78 mJy. 

The geometry between N1, C, and S1 was measured using the 2006 15 GHz data, since those components are most distinguishable. The axial ratio $N1/S1=1.59$, and the angle subtended between the arms, N1-C-S1,  is 166.9$^\circ$.

\begin{figure}
\epsscale{0.4}
\plotone{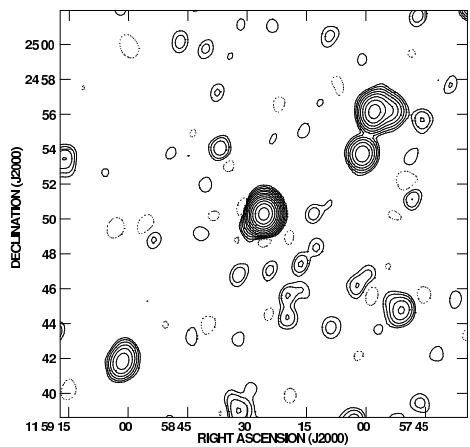}
\caption{VLA observations from March 2007 centered on J11584+2450 at 1.3649 GHz in D configuration. The contour levels begin at 0.336 mJy/beam and increase by levels of $2$.  The source shows no significant structure at this frequency and resolution, but does show possible extension to the south-east.}
\label{VLA}
\end{figure}

We used the VLA in the D-Configuration to investigate what appeared to be an extension of the western emission to kilo-parsec scale from the NRAO VLA Sky Survey (NVSS; Condon et al. 1998\nocite{1998AJ....115.1693C}), but found no indication of any western emission from J11584+2450 (Fig. \ref{VLA}) and the previous extension to be a result of the higher RMS (0.45 mJy/beam) of the NVSS compared to our image with RMS = 0.085 mJy/beam.

\begin{figure}
\epsscale{0.4}
\plotone{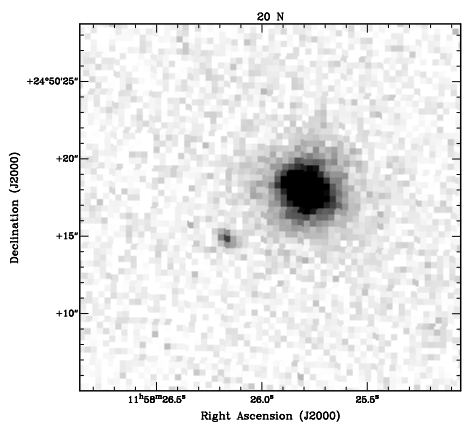}
\caption{This R-Band SDSS image of J11584+2450 shows a second source 4 arcseconds to the southeast (SDSS J115826.16+245014.9). If these sources have the same redshift, then there is 13 kpc separation between them. In this image, the magnitude of J11854+2450 is  $17.68\pm0.01$, while the magnitude of SDSS J115826.16+245014.9 is $21.99\pm0.06$.}
\label{SDSS}
\end{figure}

Additionally, we acquired a visual image (Fig. \ref{SDSS}) from the Sloan Digital Sky Survey Data Release 5 (SDSS DR5; Adelman-McCarthy et al. 2007 \nocite{2007ApJS..172..634A}). This shows the source to have a galaxy 4 arcseconds to the south east, which is uncatalogued outside of the SDSS. Two different algorithms have been used to determine a mulit-color photometric redshift for this object from the SDSS image. The first algorithm  utilized the template fitting method and yields $z=0.0008\pm0.0226$ \citep{2003AJ....125..580C} , while the second algorithm used a Neural Network method and yields $z=0.065\pm0.123$ \citep{2007arXiv0708.0030O}, placing an upper limit of $z~0.19$ which is comparable to J11584+2450's redshift.

\subsection{Spectral Index Distribution}
\label{spectral_index}

\begin{figure}
\epsscale{0.44}
\plotone{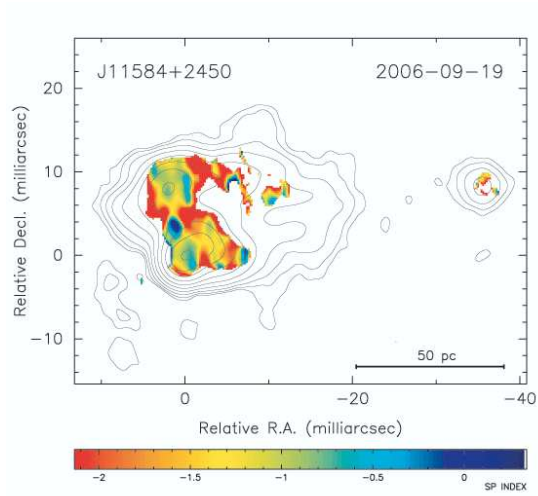}
\caption{1. Multi-Frequency observations of a newly identified CSO
(J11584+2450). 5 GHz contours overlaid on a 8 to 15 GHz
spectral index image. Notice the flat spectrum core, as well as the symmetric
dual-lobed structure in the source. Also, the emission abruptly
bends to the west. This sudden path change, and the steep spectrum
compact knot at the western edge is not clearly understood. }
\label{SpIndex}
\end{figure}

The 2006 VLBA images at 8 and 15 GHz were matched in resolution in order to obtain a spectral index distribution across the source that was overlaid onto a 5 GHz image to show overall source structure (Figure \ref{SpIndex}). This distribution clearly shows a compact flat-spectrum component ($\alpha \approx - 0.276$, where $F_\nu \propto \nu ^ \alpha$ ) situated between two steeper spectrum lobes ($\alpha \approx -1.03$). More steep spectrum emission is found to the west of these lobes ($\alpha \approx -$1.36 to $-$1.54), where it then fades below the detection threshold at high frequencies. Looking at the 5 GHz southeastern clump and using its peak, the spectral index would have to be steeper than $-$4.35 for the emission to fall below the RMS of the 8 GHz image. Alternatively, the absence of this feature could be due to having fewer short spacings at 8 GHz, or it could merely be an imaging artifact at 5 GHz.

%%%%%%%%%%%%%%%%%%%%%%%%%%%%%%%%%%%%%%%%%%%%%%%%%%%%%%%%%%%%%%%%%%%%%%%%%%%%%%%%%%%%%%%%%%%
%                               DISCUSSION                                                %
%%%%%%%%%%%%%%%%%%%%%%%%%%%%%%%%%%%%%%%%%%%%%%%%%%%%%%%%%%%%%%%%%%%%%%%%%%%%%%%%%%%%%%%%%%%

\section{Discussion}

\subsection{Reclassification of J11584+2450}

The compact flat spectrum component seen in Figure \ref{SpIndex} is compatible with emission from the nucleus, or core, of a galaxy \citep{1984RvMP...56..255B}. The steeper spectrum components extending north and south from the core are accordant with the spectral signature of jets or hot spots. This overall structure is clearly consistent with J11584+2450 being a CSO.

\subsection{Component Motions}
\label{component_motions}

To characterize intrinsic motions within the source, a multi-component elliptical gaussian model was made of the 1999 data and then applied to the 1995, 2001 and 2006 15 GHz data varying the flux and position parameters of each gaussian component (Table \ref{U_Gaussian}). The extended dual-lobed structure in the 1995 data (but notably missing from the 1999, 2001 and 2006 images) of what is now considered the core was modeled using a single component, since it is likely this extension is only an artifact.

Each of the four IFs of the 2006 data were then individually modelled and and compared to each other to determine the systematic error ($\sigma_{sys}$) associated with modelling each component. The total position error ($\sigma_{tot}$) was then calculated for individual components using $\sigma_{tot}^2=\sigma_{stat}^2+\sigma_{sys}^2$, where $\sigma_{stat}$ is the expected statistical error associated with modeling gaussian components in two dimensional polar coordinates (Table \ref{U_IFs}) (adapted from 1-dimension case per Fomalont 1999\nocite{1999ASPC..180..301F}). Since all model positions are referenced to C, $\sigma_{sys}$ for the core is accounted for by the other component uncertainties. Since the calibrated data sets from the VLBA 2cm Survery are each averaged to one frequency, the $\sigma_{sys}$ attained from the 2006 data was applied to them as well. Similarly, a multi-component gaussian model was made of the 2000 data and then applied to the 2006 8 GHz data varying the flux and position parameters of each gaussian component (Table \ref{X_Gaussian}). $\sigma_{sys}$ values were obtained for each component using the four IFs as above in the 2006 data, and using the four IFs closest to those same frequencies in the 2000 data (Table \ref{X_IFs}).

\begin{figure}
\epsscale{0.4}
\plotone{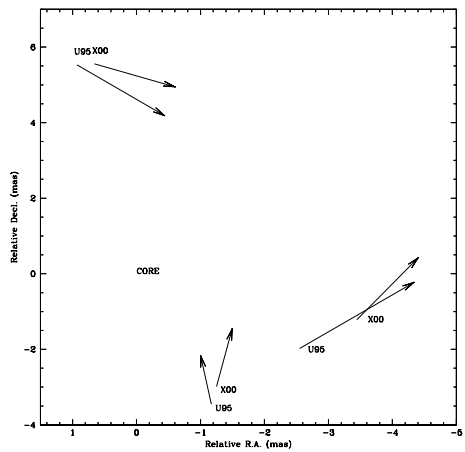}
\caption{Relative velocity of model components. Velocity of each gaussian model component is plotted ($1  \rm{mas} = 0.2c$) with the tail of each vector originating at the model componentÕs position at its earliest observation (1995 for 15 GHz \& 2000 for 8 GHz). The model
fits are in agreement with the contour
overlay plots in showing this
source to be shrinking.}
\label{motions}
\end{figure}

\begin{figure}
\epsscale{0.44}
\plotone{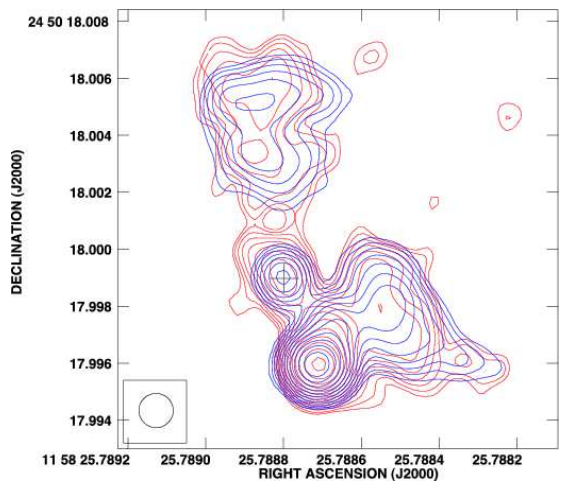}
\caption{Two epoch overlay of J11584+2450. 15 GHz data from 1995 (Red Contours) is plotted with the 15 GHz data from 2006 (Blue Contours). This figure shows contraction of the source towards the core (denoted by a cross) over time. The contour levels begin at 1.10 mJy/beam and increase by factors of $2^{1/2}$. }
\label{U9506O}
\end{figure}

These models were used to calculate component velocities relative to the core for each band, which are plotted in Fig. \ref{motions} with the tail of each vector located at the earliest data position in the band. The components representing the hot spots at the working surface of the jets which were modeled (N1 and S1) appear to have contracted towards the core as well as traveled westward, and this motion is consistent between the two bands. Additionally, the W2 component moves towards the northwest in both bands. Performing a least-squares fit to solve for the velocities in each band separately, and then using these independent values to reduce the error yields a radial contraction velocity (normalized to the speed of light) of $0.42 \pm 0.03$ c for S1 and $0.20 \pm 0.07$ c for N1. Overlaying the contour maps of different epochs (see Fig. \ref{U9506O} for one example) is also supportive of contraction, since the 1995 contours are interior to the 2006 contours suggesting that the component motion is not an artifact of the modeling.

This motion was not detected by the VLBA 2 cm Survey over the six year interval between 1995 and 2001 since the velocities involved ($\sim0.03$ mas/year) are well within most of their stated velocity errors for this source \citep{2004ApJ...609..539K}.
The total flux density of the source has been decreasing steadily since at least 1995, the models show that this drop can be attributed to S1 decreasing steadily (40\% decrease at 15 GHz over this 11 year period), while the other components exhibit small fluctuations.

\subsection{Apparent Motion Interpretation}

Since actual contraction of the source towards the core is something that has not been previously observed, we first examine reasons behind an effect that would merely cause apparent motion in the system. One possible explanation for seeing the contraction of this source is that if hotspots are advancing out away from the core and expanding and younger hotspots are brightening due to interactions at the end of  the jet, then the models might not be fit to the same components. The largest problems with this hypothesis are its lack of explanation of both the western emission and the western component to the hotspot velocities, which means these properties require a separate unrelated explanation if the contraction is to be explained by hot spot dimming and advance. 

\subsection{Physical Motion Interpretations}
 
Leaving open the possibility that the data represent physical motions in the system, we include discussion along those lines. One interpretation is that we are viewing a projection effect caused by rotation of the source. While solid body rotation is an unphysical scenario, it gives us an idea about what fluid rotation would look like for this system so we consider it as a first approximation of rotational motion. The angular velocity of the rotation is dependent upon the initial orientation. Assuming the axis of rotation lies in the plane of the sky, and the inclination angle is less than 45$^\circ$ since larger values would yield Doppler boosting, the jets would have a rotational period between just 260 and 1880 years. 

Since the system is a fluid and not a rigid body it would actually have differential rotation.  The core would therefore be spinning even faster than the observed jet components and we would expect the core to appear highly variable since the base of the jet would frequently be pointed towards us. A second physical interpretation of the motion is precession. The westward component to the velocities of both jets strongly argues against precession. If the jets were precessing and thus appearing shorter, they would move in opposite directions (i.e. one moves east while the other moves west) as well as inward. Additionally, these hypotheses fail to explain the older western emission. 

Another physical elucidation of the motion is that there exists some reason for the pressure of the environment to increase, then this could leave the jets under-pressured and lead to contraction of J11584+2450. Such a pressure change could result from a relative motion between a clumpy environment and the CSO. The 1.59 $N1/S1$ axial ratio is also indicative of the jets encountering a dense environment. If the departure of the ratio from 1 was due to Doppler boosting then the brighter hotspot would be further from the core (S1 in the case of J11584+2450). This would yield a ratio smaller than 1, therefore the ratio is likely due to the jet running into difficulty as it tunnels through the environment causing it to be both shorter and brighter.  Both the angle N1-C-S1  and the western deviation of the emission is consistent with the source moving eastward and being influenced by ram pressure similar to what has been observed in wide-angle tailed radio sources (e.g. 3C465, Hardcastle et al. 2005\nocite{2005MNRAS.359.1007H}; Sakelliou \& Merrifield 2000\nocite{2000MNRAS.311..649S}), but on a smaller spacial scale. Relative motion could exist between J11584+2450 and its host galaxy causing the interstellar medium to produce  ram pressure against the radio jet. If the companion galaxy to the south east (Fig. \ref{SDSS}) is at a similar redshift and these two galaxies are members of a cluster then J11584+2450 might be moving towards the center of the gravitational potential. However, this is highly speculative and more observations are needed to determine the cluster environment.

%This relative motion could possibly be due to the companion galaxy to the south east (Fig. \ref{SDSS}) assuming the two galaxies are at a similar redshift.

%%%%%%%%%%%%%%%%%%%%%%%%%%%%%%%%%%%%%%%%%%%%%%%%%%%%%%%%%%%%%%%%%%%%%%%%%%%%%%%%%%%%%%%%%%%
%                               CONCLUSIONS                                                %
%%%%%%%%%%%%%%%%%%%%%%%%%%%%%%%%%%%%%%%%%%%%%%%%%%%%%%%%%%%%%%%%%%%%%%%%%%%%%%%%%%%%%%%%%%%

\section {Conclusions}

After analyzing multifrequency (5, 8, and 15 GHz) VLBA data from radio source J11584+2450 we reclassify it as a CSO. Fitting the data with multi-component gaussian models and overlaying images of different epochs on each other not only show that this source is not growing at the usual rate of $\sim$ 0.1 to 0.3 c, but each jet is apparently shrinking in size at $\sim$  0.3 c and each is additionally moving westward at $\sim$ 0.2 c.  Confirmation of other CSOs having either recessive behavior or non-radial motion like 1031+567 \citep{2000ApJ...541..112T} would mean the current models need to be modified to allow for possible non-linear growth periods during the evolution of AGN. The prospect of non-linear growth for CSOs would further bring in to question the validity of kinematic ages. \citet{2005ApJ...622..136G} found 7 out of the 13 CSO they dated to be under 500 yr old and commented that the expectation for a steady-state population of CSOs would have a uniform distribution of ages.  While this is a small statistical sample it is also what one would expect to see if CSOs spend a greater fraction of time as small sources. However, in kinematic observations of $\sim$ 10 CSOs we have seen contraction in just 1 source.

Future VLBA observations of this source are planned to follow the motion of the components and to see whether they continue to recede towards the core, and what time-scale this occurs over. Lower frequency (1.4 GHz) observations should be carried out to confirm the existence of the 5 GHz southeastern clump. If the emission from both jets is actually flowing towards the west, lower frequency observations might also show W1 merging with the diffuse western emission, and could reveal larger scale structures.

\begin{acknowledgements}
We thank an anonymous referee for constructive suggestions. The National Radio Astronomy Observatory is a facility of the National Science Foundation operated under cooperative agreement by Associated Universities, Inc.
\end{acknowledgements}

{\it Facilities:} \facility{VLA ()}, \facility{VLBA ()}

%%%%%%%%%%%%%%%%%%%%%%%%%%%%%%%%%%%%%%%%%%%%%%%%%%%%%%%%%%%%%%%%%%%%%%%%%%%%%%%%%%%%%%%%%%%
%                               REFERENCES                                                %
%%%%%%%%%%%%%%%%%%%%%%%%%%%%%%%%%%%%%%%%%%%%%%%%%%%
%%%%%%%%%%%%%%%%%%%%%%%%%%%%%%%%%%%%%%%%
\bibliographystyle{apj}
\bibliography{references}

\clearpage

%%%%%%%%%%%%%%%%%%%%%%%%%%%%%%%%%%%%%%%%%%%%%%%%%%%%%%%%%%%%%%%%%%%%%%%%%%%%%%%%%%%%%%%%%%%
%                               FIGURES                                                   %
%%%%%%%%%%%%%%%%%%%%%%%%%%%%%%%%%%%%%%%%%%%%%%%%%%%%%%%%%%%%%%%%%%%%%%%%%%%%%%%%%%%%%%%%%%%

%%%%%%%%%%%%%%%%%%%%%%%%%%%%%%%%%%%%%%%%%%%%%%%%%%%%%%%%%%%%%%%%%%%%%%%%%%%%%%%%%%%%%%%%%%%
%                               TABLES                                                    %
%%%%%%%%%%%%%%%%%%%%%%%%%%%%%%%%%%%%%%%%%%%%%%%%%%%%%%%%%%%%%%%%%%%%%%%%%%%%%%%%%%%%%%%%%%%

\begin{deluxetable}{lccccccc}
\tabletypesize{\scriptsize}
\tablecolumns{8}
\tablewidth{0pt}
\tablecaption{VLA and VLBA Observations of J1158+2450.\label{Observations}}
\tablehead{\colhead{Freq.}&\colhead{Date}&\colhead{Time}&\colhead{BW}
&\colhead{Pol.}&\colhead{IFs}&\colhead{Peak}&\colhead{rms}
 \\
\colhead{(GHz)} & \colhead{} & \colhead{(minutes)} & \colhead{(MHz)} 
& \colhead{}&\colhead{}&\colhead{(mJy beam$^{-1}$)}&\colhead{(mJy beam$^{-1}$)}}

\startdata
1.3649\tablenotemark{*} & 2007 Mar 08 & 24.7 & 100 & 4 & 2 & 1073.7 & 0.1 \\
4.8447\tablenotemark{\dagger} & 2006 Sep 19 & 24.9 & 32 &4 & 4 & 193.38 & 0.11 \\
8.3447\tablenotemark{\dagger} & 2006 Sep 19 &  26.9 & 32 & 4 & 4 & 118.38 & 0.17 \\
8.3541\tablenotemark{\dagger} & 2000 May 06 & 20.6 & 64 & 1 & 8 & 133.1 & 0.2 \\
15.138\tablenotemark{\dagger} & 2006 Sep 19 & 90.8 & 32 & 4 & 4 & 52.27 &  0.11\\
15.335\tablenotemark{\dagger} \tablenotemark{\ddagger}&  2001 Mar 04 & 57.0 & 56 & 1 & 1 & 74.85 & 0.31\\
15.335\tablenotemark{\dagger} \tablenotemark{\ddagger}& 1999 May 21 & 37.6 & 56 &1 &1 & 82.73 & 0.33 \\
15.350\tablenotemark{\dagger}\tablenotemark{\ddagger} & 1995 Apr 07 & 44.9 & 32 &1 &1 & 110.0 & 0.3 \\

\enddata

\tablenotetext{*}{VLA Observation}
\tablenotetext{\dagger}{VLBA Observation}
\tablenotetext{\ddagger}{These data were taken as part of the VLBA 2 cm Survery, \cite{2004ApJ...609..539K}}

\end{deluxetable}

\begin{deluxetable}{lccccccccc}
\tabletypesize{\scriptsize}
\tablecolumns{10}
\tablewidth{0pt}
\tablecaption{15 GHz Gaussian Model Components\tablenotemark{*}.\label{U_Gaussian}}
\tablehead{\colhead{Component}&\colhead{Epoch}&\colhead{$S$}&\colhead{$r$}
&\colhead{$\sigma_r$}&\colhead{$\theta$}&\colhead{$\sigma_\theta$}&\colhead{$a$}&\colhead{$b/a$}&\colhead{$\Phi$}
\\
\colhead{} & \colhead{} & \colhead{(Jy)} & \colhead{(mas)}& \colhead{(mas)} 
& \colhead{($^o$)}& \colhead{($^o$)}&\colhead{(mas)}&\colhead{}&\colhead{($^o$)}}
\startdata
C... & 1995 & 0.0167 & 0.000 & 0.007 & 0.00 & 0.00 & 0.283 & 1.00 & -18.57 \\
       & 1999 & 0.0331 & 0.000 & 0.003 & 0.00 & 0.00 & 0.283 & 1.00 & -18.57 \\
       & 2001 & 0.0170 & 0.000 & 0.007 & 0.00 & 0.00 & 0.283 & 1.00 & -18.57\\
       & 2006 & 0.0120 & 0.000 & 0.003 & 0.00 & 0.00 & 0.283 & 1.00 & -18.57\\
S1...&1995 & 0.1267 & 3.647 & 0.014 & -161.28 & 0.15 & 0.272 & 0.86 & 50.59\\
        &1999 & 0.0945 & 3.514 & 0.014 & -159.79 & 0.15 & 0.272 & 0.86 & 50.59\\
        & 2001 & 0.0798 & 3.400 & 0.014 & -159.03 & 0.15 &  0.272 & 0.86 & 50.59\\
        &2006 & 0.0765 & 3.360 & 0.014 & -160.12 & 0.15 & 0.272 & 0.86 & 50.59\\
N1...&1995 & 0.0327 & 5.610 & 0.052 & 9.55 & 1.12 & 3.02 & 0.345 & 29.62\\
        &1999 & 0.0288 & 5.801 & 0.049 & 8.37 & 0.95 & 3.02 & 0.345 & 29.62 \\
        & 2001 & 0.0257 & 5.497 & 0.045 & 9.52 & 1.78 & 3.02 & 0.345 & 29.62 \\
        &2006 & 0.0399 & 5.325 & 0.038 & 6.82 & 0.50 & 3.02 & 0.345 & 29.62\\
W2...&1995 & 0.0637 & 3.226 & 0.034 & -127.77 & 0.46 & 2.72 & 0.32 & -81.22\\
        &1999 & 0.0541 & 3.775 & 0.033 & -115.06 & 0.46 & 2.72 & 0.32 & -81.22\\
        & 2001 & 0.0369 & 3.746 & 0.033 & -115.65 & 2.65 & 2.72 & 0.32 & -81.22\\
        &2006 & 0.0630 & 3.435 & 0.030 & -116.83 & 0.34 & 2.72 & 0.32 & -81.22\\

\enddata
\tablenotetext{*}{NOTE - Parameters of each Gaussian component of the model
brightness distribution are as follows:  Component, Gaussian component; 
Epoch, year of observation (see Table \ref{Observations});  
$S$, flux density;  $r$, $\sigma_r$, $\theta$, $\sigma_\theta$, polar coordinates (and the associated errors) of the center of the component relative to the center of component C; $a$, semimajor axis; $b/a$, axial ratio; 
$\Phi$, component orientation. 
All angles are measured
from north through east. }
\end{deluxetable}

\begin{deluxetable}{lccccccccccc}
\tabletypesize{\scriptsize}
\tablecolumns{12}
\tablewidth{0pt}
\tablecaption{15 GHz Gaussian Model Components from Selected IFs\tablenotemark{*}.\label{U_IFs}}
\tablehead{\colhead{Component}&\colhead{Epoch}&\colhead{$r_1$}&\colhead{$r_2$}
&\colhead{$r_3$}&\colhead{$r_4$}&\colhead{$\sigma_{r(sys)}$}&\colhead{$\theta_1$}&\colhead{$\theta_2$}&\colhead{$\theta_3i$}&\colhead{$\theta_4$}&\colhead{$\sigma_{\theta(sys)}$}
\\
\colhead{} & \colhead{} & \colhead{(mas)} & \colhead{(mas)}& \colhead{(mas)} 
& \colhead{(mas)}& \colhead{(mas)}&\colhead{($^o$)}&\colhead{($^o$)}&\colhead{($^o$)}&\colhead{($^o$)}&\colhead{($^o$)}}
\startdata
S1...&2006 & 3.368 & 3.353 & 3.341 & 3.377 & 0.014 & -159.93 & -160.10 & -160.36 & -160.13 & 0.15\\
N1...&2006 & 5.276 & 5.343 & 5.376	 & 5.318 & 0.037 & 7.20 & 6.19 & 6.75 & 7.19 & 0.41\\
W2...&2006 & 3.473 & 3.430 & 3.390 & 3.440 & 0.030 & -116.45 & -116.63 & -117.06 & -117.30 & 0.34\\

\enddata
\tablenotetext{*}{NOTE - Parameters of each Gaussian component of the IF
model position distribution are as follows:  Component, Gaussian component; 
Epoch, year of observation (see Table \ref{Observations});  
$r_1-r_4$, radial positions of model components in IFs 1 through 4 respectively; $\sigma_{r(sys)}$, the standard deviation in radial position; $\theta_1-\theta_4$, the polar angular position of model components in IFs 1 through 4 respectively; $\sigma_{\theta(sys)}$, the standard deviation in polar angular position. All angles are measured from north through east.}
\end{deluxetable}

\begin{deluxetable}{lccccccccc}
\tabletypesize{\scriptsize}
\tablecolumns{10}
\tablewidth{0pt}
\tablecaption{8 GHz Gaussian Model Components\tablenotemark{*}.\label{X_Gaussian}}
\tablehead{\colhead{Component}&\colhead{Epoch}&\colhead{$S$}&\colhead{$r$}
&\colhead{$\sigma_r$}&\colhead{$\theta$}&\colhead{$\sigma_\theta$}&\colhead{$a$}&\colhead{$b/a$}&\colhead{$\Phi$}
\\
\colhead{} & \colhead{} & \colhead{(Jy)} & \colhead{(mas)}& \colhead{(mas)} 
& \colhead{($^o$)}& \colhead{($^o$)}&\colhead{(mas)}&\colhead{}&\colhead{($^o$)}}
\startdata
C...  & 2000 & 0.0273 & 0.000 & 0.001 & 0.00 & 0.00  & 0.214 & 1.00 & 63.74\\
        & 2006 & 0.0115 & 0.000 & 0.003 & 0.00 & 0.00 & 0.214 & 1.00 & 63.74\\
S1...& 2000 & 0.1581 & 3.240 &  0.017 & -157.29 & 0.15 & 0.4875& 0.71 & -23.38\\
        & 2006 & 0.1507 & 3.082 & 0.022 & -155.45 & 0.19 & 0.4875 & 0.71 & -23.38\\
N1...& 2000 & 0.1005 & 5.597 & 0.055 & 6.75 & 0.31 & 2.195 & 0.76 & -38.51\\
        & 2006 & 0.1098 & 5.508 & 0.049 & 5.27 & 0.10 & 2.195 & 0.76 & -38.51\\
W2...& 2000 & 0.1505 & 3.653 & 0.028 & -109.43 & 0.32  & 1.57 & 0.80 & 40.08\\
         & 2006 & 0.1489 & 3.703 & 0.011 & -105.95 & 0.47 & 1.57 & 0.80 & 40.08\\

\enddata
\tablenotetext{*}{NOTE - Parameters of each Gaussian component of the model
brightness distribution are as follows:  Component, Gaussian component; 
Epoch, year of observation (see Table \ref{Observations});  
$S$, flux density; $r$, $\sigma_r$, $\theta$, $\sigma_\theta$, polar coordinates (and the associated errors) of the center of the component relative to the center of component C; $a$, semimajor axis; $b/a$, axial ratio; 
$\Phi$, component orientation. 
All angles are measured
from north through east.}
\end{deluxetable}

\begin{deluxetable}{lccccccccccc}
\tabletypesize{\scriptsize}
\tablecolumns{12}
\tablewidth{0pt}
\tablecaption{8 GHz Gaussian Model Components from Selected IFs\tablenotemark{*}.\label{X_IFs}}
\tablehead{\colhead{Component}&\colhead{Epoch}&\colhead{$r_1$}&\colhead{$r_2$}
&\colhead{$r_3$}&\colhead{$r_4$}&\colhead{$\sigma_{r(sys)}$}&\colhead{$\theta_1$}&\colhead{$\theta_2$}&\colhead{$\theta_3i$}&\colhead{$\theta_4$}&\colhead{$\sigma_{\theta(sys)}$}
\\
\colhead{} & \colhead{} & \colhead{(mas)} & \colhead{(mas)}& \colhead{(mas)} 
& \colhead{(mas)}& \colhead{(mas)}&\colhead{($^o$)}&\colhead{($^o$)}&\colhead{($^o$)}&\colhead{($^o$)}&\colhead{($^o$)}}
\startdata
S1...&2000 & 3.225 & 3.270 & 3.250 &  3.237 & 0.017 & -157.64 & -157.48 & -157.26 & -157.29 & 0.15\\
        &2006 & 3.106 & 3.079 & 3.051 & 3.101 & 0.022 & -155.60 & -155.27 & -155.24 & -155.67 & 0.19\\
N1...&2000 & 5.573 & 5.675 & 5.535 & 5.551 & 0.055 & 7.31 & 6.65 & 7.05 & 6.54 & 0.31\\
        &2006 & 5.467 & 5.501 & 5.586 & 5.467 & 0.049 & 5.17 & 5.34 & 5.19 & 5.39 & 0.09\\
W2...&2000 & 3.607 & 3.673 & 3.675 & 3.642 & 0.028 & -109.24 & -109.98 & -110.07 & -109.45 & 0.32\\
        &2006 & 3.692 & 3.721 & 3.696 & 3.707 & 0.011 & -106.46 & -105.96 & -105.23 & -106.29 & 0.47\\
\enddata
\tablenotetext{*}{NOTE - Parameters of each Gaussian component of the IF
model position distribution are as follows:  Component, Gaussian component; 
Epoch, year of observation (see Table \ref{Observations});  
$r_1-r_4$, radial positions of model components in IFs 1 through 4 respectively; $\sigma_{r(sys)}$, the standard deviation in radial position; $\theta_1-\theta_4$, the polar angular position of model components in IFs 1 through 4 respectively; $\sigma_{\theta(sys)}$, the standard deviation in polar angular position. All angles are measured from north through east.}
\end{deluxetable}

\end{document}